\documentclass[aps,preprint,nofootinbib,floatfix]{revtex4}
\usepackage{amssymb}
\usepackage{graphicx}
\usepackage{amsmath}

\graphicspath{{./img/}}

\input{tcilatex}

\begin{document}

\title{Generalized On-Shell Ward Identities in String Theory}
\author{Jen-Chi Lee}
\affiliation{Institute of Physics, National Chiao-Tung University, Hsin Chu, Taiwan, ROC}
\date{\today}

\begin{abstract}
It is demonstrated that an infinite set of string-tree level on-shell Ward
identities, which are valid to all $\sigma $-model loop orders, can be
systematically constructed without referring to the string field theory. As
examples, bosonic massive scattering amplitudes are calculated explicitly up
to the second massive excited states. Ward identities satisfied by these
amplitudees are derived by using zero-norm states in the spetrum. In
particular, the inter-particle Ward identity generated by the $D_{2}\otimes
D_{2}^{\prime }$ zero-norm state at the second massive level is
demonstrated. The four physical propagating states of this mass level are
then shown to form a large gauge multiplet. This result justifies our
previous consideration on higher inter-spin symmetry from the generalized
worldsheet $\sigma $-model point of view.
\end{abstract}

\maketitle

%
%-----------------------------------------------------------------------------

%-----------------------------------------------------------------------------

Since Veneziano \cite{1} derived the massless gravitational and Yang-Mills
Ward identities for hidden symmetries of string theories from the canonical
transformations of the string phase-space path integral, there has been an
attempt by Kubota \cite{2} to generate new Ward identities corresponding to
higher massive string states. Recently, the target space-time $w_{\infty }$%
-symmetries of 2D string theory, first proposed by Avan and Jevicki \cite{3}
in the context of the collective field representation of $c=1$ matrix model 
\cite{4}, where suggested to associate with higher-level string states. \cite%
{5} Subsequently, the corresponding on-shell Ward identities were discussed
by Klebanov and many authors. \cite{6}

The proposal of Kubota and the encouraging results of the toy 2D string
theory as discussed above suggest that there exist an infinite set of Ward
identities in the higher dimensional critical string theories. These Ward
identities are expected to be associated with string symmetries proposed in
Ref. \cite{7} where symmetries are directly related to the zero-norn states
in the spectrum, and may even be related to the works of Gross \cite{8} and
Atick and Witten \cite{9}, which claim infinite symmetry structure of string
gets restored at very high energy. However, the approach of Kubota uses the
vertex operator proposed by Ichinose and Sakita. \cite{10} The vertex
operator of Ref. \cite{10} contains a string field in it, hence is analogous
to the unpleasant second-quantized string field theory formalism. In
addition, there are other drawbacks in the above approach, e.g., the
off-shell ambiguities remain and there seems to have no underlying principle
to choose appropriate gauge functions which generate the proposed higher
spin Ward identities. These drawbacks make it difficult to make any physical
interpretations of the identities (see Ref. \cite{7}). A general principle
which allows one to explicitly write down these Ward identities seems
necessary.

In this paper, we will derive the explicit form of massive on-shell Ward
identities by using zero-norm states in the bosonic string spectrum. The
advantages of our approach are that off-shell ambiguities are avoided, and
we can easily write down an infinite set of string-tree level on-shell Ward
identities after a systematic construction of zero-norm states in the
spectrum. These Ward identities, which are valid to $all$ $\sigma $-model
loop $(\alpha ^{\prime })$ orders, indicate that quantum string theories do
possess an infinite number of high energy symmetry structures as was
conjectured in Ref. \cite{7}-\cite{9}. We will first calculate the massive
string-tree level scattering amplitudes up to the second massive states to
derive the corresponding Ward identities. Of particular interest, the Ward
identity corresponding to the $D_{2}\otimes D_{2}^{\prime }$ zero-norn state 
\cite{11} explicitly relates amplitudes of $four$ $different$ $spin$ $states$
at the second massive level. The corresponding symmetry transformation law (
in the first order weak field approximation ) of the four background fields
can then be constructed. This result justifies our previous consideration on
higher inter-spin symmetries from the generalized $\sigma $-model point of
view, and is a general feature for higher massive level. Although the
decoupling of zero-norm states from the string amplitudes has been proved
for a long time by the so-called \textquotedblleft $canceled$ $propagator$ $%
argument$\textquotedblright in the context of \textquotedblleft old
fashioned\textquotedblright operator method, \cite{12} its implication on
the physical amplitudes was always ignored and not clear so far. On the
other hand, our approach will be based on Polyakov functional integral
method. Moreover, through the use of explicit form of zero-norm states which
can be calculated to any higher massive level, one can easily write down
infinite relations among string scattering amplitudes. Our results can be
generalized to the superstring massive states. One can easily write down the
symmetry transformation laws of the massive background fields after
explicitly deriving the Ward identities. These transformation laws turn out
to be too messy to obtain from the worldsheet $\sigma $-model approach. \cite%
{13}.

For simplicity we will consider one excited state whose decay process going
into three tachyons, and we will just list amplitudes for the $s-t$ channel
only throughout the paper. The open string massless vector amplitude has
been calculated \cite{14} to be ( we use the notation of Ref.\cite{14} and
assume product of fields at the same point to be normal ordered throughout
the text )%
\begin{equation}
A=\int \overset{4}{\underset{i=1}{\prod }}dx_{i}<e^{ik_{1}\cdot
X(x_{1})}\,\varepsilon \cdot \partial X(x_{2})\,e^{ik_{2}\cdot
X(x_{2})}e^{ik_{3}\cdot X(x_{3})}e^{ik_{4}\cdot X(x_{4})}>=\varepsilon _{\mu
}T_{0}^{\mu }  \label{1a}
\end{equation}%
with%
\begin{equation}
T_{0}^{\mu }=\frac{\Gamma (-\frac{s}{2}-1)\Gamma (-\frac{t}{2}-1)}{\Gamma (%
\frac{u}{2}+2)}[k_{3}^{\mu }(s/2+1)-k_{1}^{\mu }(t/2+1)],  \label{2a}
\end{equation}%
where $\varepsilon _{\mu }$ is the vector polarization and $s,t$ \ and $\mu $
are the usual Mandelstam variables,%
\begin{equation}
s=-(k_{1}+k_{2})^{2},t=-(k_{2}+k_{3})^{2},u=-(k_{1}+k_{3})^{2},s+t+u=\sum
(mass)^{2}.  \label{3a}
\end{equation}%
There is only one singlet massless zero-norm state \cite{7} in the spectrum,
which is 
\begin{equation}
k\cdot \alpha _{-1}|0,k>,-k^{2}=m^{2}=0.  \label{4a}
\end{equation}%
The corresponding Ward identity is easily checked to be%
\begin{equation}
k_{2\mu }T_{0}^{\mu }=[(k_{2}\cdot k_{3})(s/2+1)-(k_{2}\cdot k_{1})(t/2+1)=0.
\label{5a}
\end{equation}%
This is consistent with the result obtained by Veneziano through the
canonical transformation of the string variables. Note that the identity in
Eq.(\ref{5a}) is valid to all $\sigma $-model loop $(\alpha ^{\prime })$
orders.

We now turn to the first massive level state. There is only one
positive-norm physical propagating mode at this level. The most general form
of polarization is given by%
\begin{equation}
(\varepsilon _{\mu \nu }\alpha _{-1}^{\mu }\alpha _{-1}^{\nu }+\varepsilon
_{\mu }\alpha _{-2}^{\mu })|0,k>,\varepsilon _{\mu \nu }\equiv \varepsilon
_{\nu \mu }  \label{6a}
\end{equation}%
with gauge conditions%
\begin{equation}
\varepsilon _{\nu }=-k^{\mu }\varepsilon _{\mu \nu },\varepsilon _{\mu
}^{\mu }-2k^{\mu }k^{\nu }\varepsilon _{\mu \nu }=0,k^{2}=-2.  \label{7a}
\end{equation}%
The amplitude is defined to be%
\begin{equation}
T_{1}^{\mu \nu }=\int \overset{4}{\underset{i=1}{\prod }}dx_{i}<e^{ik_{1}%
\cdot X}\partial X^{\mu }(x_{2})\,\partial X^{\nu }(x_{2})e^{ik_{2}\cdot
X}e^{ik_{3}\cdot X}e^{ik_{4}\cdot X}>,  \label{8a}
\end{equation}%
\begin{equation}
T_{1}^{\mu }=\int \overset{4}{\underset{i=1}{\prod }}dx_{i}<e^{ik_{1}\cdot
X}\partial ^{2}X^{\mu }(x_{2})\,e^{ik_{2}\cdot X}e^{ik_{3}\cdot
X}e^{ik_{4}\cdot X}>.  \label{9a}
\end{equation}%
The polarization doublet $(\varepsilon _{\mu \nu },\varepsilon _{\mu })$ and
the amplitude doublet $(T_{1}^{\mu \nu },T_{1}^{\mu })$ as given above
describe the dynamics of the first massive state. We will use the method
suggested in Ref.\cite{14} to calculate Eqs.(\ref{8a}) and (\ref{9a}). To
calculate Eq.(\ref{8a}), we first absorb the kinematic factor to the
exponent,%
\begin{equation}
\varepsilon _{\mu \nu }\partial X^{\mu }\,\partial X^{\nu }e^{ik\cdot
X}\rightarrow \exp [ik\cdot X+i\varepsilon ^{(1)}\cdot \partial
X+i\varepsilon ^{(2)}\cdot \partial X],  \label{10a}
\end{equation}%
then the correlation function can be obtained by using the following formula:%
\cite{11}%
\begin{equation}
<:e^{A}::e^{A}:...:e^{A}:>=\exp [\underset{i<j}{\sum }<A_{i}A_{i}>].
\label{11a}
\end{equation}%
It is understood that only terms multilinear in $\varepsilon ^{(1)}$ and $%
\varepsilon ^{(2)}$ are picked up after evaluating the correlation function,
and the factors $\varepsilon _{\mu }^{(1)}\varepsilon _{\nu }^{(2)}$ should
be replaced by $\varepsilon _{\mu \nu }$. This method can be generalized to
any higher rank polarization with arbitrary higher derivative $\partial
^{n}X^{\mu }$ in the vertex operator. To make the $SL(2,R)\,$\ gauge fixing
which corresponds to the well-known M\"{o}bius transformation, we choose $%
x_{1}=0,0\leq x_{2}\leq 1,x_{3}=1,x_{4}=\infty $. After some calculation, we
get%
\begin{eqnarray}
T_{1}^{\mu \nu } &=&\frac{\Gamma (-\frac{s}{2}-1)\Gamma (-\frac{t}{2}-1)}{%
\Gamma (\frac{u}{2}+2)}  \notag \\
&&\times \lbrack s/2(s/2+1)k_{3}^{\mu }k_{3}^{\nu }+t/2(t/2+1)k_{1}^{\mu
}k_{1}^{\nu }-2(s/2+1)(t/2+1)k_{1}^{\mu }k_{3}^{\nu }],  \label{12a}
\end{eqnarray}%
\begin{equation}
T_{1}^{\mu }=\frac{\Gamma (-\frac{s}{2}-1)\Gamma (-\frac{t}{2}-1)}{\Gamma (%
\frac{u}{2}+2)}[-k_{3}^{\mu }s/2(s/2+1)-k_{1}^{\mu }t/2(t/2+1)].  \label{13a}
\end{equation}

There are two types of zero-norm states in the bosonic open string spectrum. 
\cite{7} They can be obtained either by standard spectrum analysis or
generated in the following way,

Type I%
\begin{equation}
L_{-1}|\chi >\text{ where }L_{m}|\chi >=0,\text{ }m\geq 1,\text{ }L_{0}|\chi
>=0,  \label{14a}
\end{equation}

Type II%
\begin{equation}
(L_{-2}+3/2L_{-1}^{2})|\overset{\sim }{\chi }>,\text{ where }L_{m}|\overset{%
\sim }{\chi }>=0,\text{ }m\geq 1,\text{ }(L_{0}+1)|\overset{\sim }{\chi }>=0.
\label{15a}
\end{equation}%
Type I states have zero norm at any space time dimension, whereas type II
states have zero norm only at $D=26$. A systematic construction of these
zero-norm states has been given in Ref.\cite{7}. At the first massive level,
we have a vector type I zero-norm state%
\begin{equation}
\lbrack (\theta \cdot \alpha _{-1})(k\cdot \alpha _{-1})+\theta \cdot \alpha
_{-2}]|0,k>,\text{ }\theta \cdot k=0,\text{ }-k^{2}=m^{2}=2,  \label{16aa}
\end{equation}%
and a singlet type II zero-norm state%
\begin{equation}
\lbrack 1/2\alpha _{-1}\cdot \alpha _{-1}+3/2(k\cdot \alpha
_{-1})^{2}+5/2k\cdot \alpha _{-2}]|0,k>.  \label{17a}
\end{equation}%
The corresponding Ward identities are%
\begin{equation}
k_{(\mu }\theta _{\nu )}T_{1}^{\mu }+\theta _{\mu }T_{1}^{\mu }=\frac{\Gamma
(-\frac{s}{2}-1)\Gamma (-\frac{t}{2}-1)}{\Gamma (\frac{u}{2}+2)}P(s,t)=0
\label{18a}
\end{equation}%
and 
\begin{equation}
(3/2k_{\mu }k_{\nu }+1/2\eta _{\mu \nu })T_{1}^{\mu \nu }+5/2k_{\mu
}T_{1}^{\mu }=\frac{\Gamma (-\frac{s}{2}-1)\Gamma (-\frac{t}{2}-1)}{\Gamma (%
\frac{u}{2}+2)}P^{\prime }(s,t)=0,  \label{19a}
\end{equation}%
where we have used the on-shell condition of $k_{i}$. The factors $P$ \ and $%
P^{\prime }$ are polynomials of $s$ and $t$, and can be explicitly verified
to be zero after some calculation. Equations (\ref{18a}) and (\ref{19a}) are
Ward identities associated with the stringy symmetries derived from
worldsheet $\sigma $-model point of view in Ref. \cite{7}.

We now come to deriving the interesting inter-particle Ward identity which
begins to show up at the second massive states. We will first discuss the
open string case. Before doing this, let us first classify the physical
propagating states at this mass level.\cite{15} The most general form of the
first state $\Psi _{1}$ is given by%
\begin{equation}
\{\varepsilon _{\mu \nu \lambda }\alpha _{-1}^{\mu }\alpha _{-1}^{\nu
}\alpha _{-1}^{\lambda }+\varepsilon _{(\mu \nu )}\alpha _{-1}^{(\mu }\alpha
_{-2}^{\nu )}+\varepsilon _{\mu }\alpha _{-3}^{\mu }\}|0,k>,\varepsilon
_{\mu \nu \lambda }=\varepsilon _{(\mu \nu \lambda )}  \label{20a}
\end{equation}%
with gauge conditions%
\begin{equation*}
\varepsilon _{(\mu \nu )}=-3/2k^{\lambda }\varepsilon _{\mu \nu \lambda
},\varepsilon _{\mu }=1/2k^{\nu }k^{\lambda }\varepsilon _{\mu \nu \lambda
},-k^{2}=4
\end{equation*}%
\begin{equation}
2\varepsilon _{\mu \lambda }^{\mu }-k^{\mu }k^{\nu }\varepsilon _{\mu \nu
\lambda }=0.  \label{21aa}
\end{equation}%
Note that the form of Eqs.(\ref{20a}) and (\ref{21aa}) differs from those of
Ref. \cite{15} by zero-norm states. One might call $\Psi _{1}$ the totally
symmetric\textquotedblleft spin\textquotedblright three state, and the
remaining polarization $\varepsilon _{(\mu \nu )}$ and $\varepsilon _{\mu }$
are mere gauge artifacts.

The amplitudes are given by%
\begin{equation}
T_{2}^{\mu \nu \lambda }=\int \overset{4}{\underset{i=1}{\prod }}%
dx_{i}<e^{ik_{1}\cdot X}\partial X^{\mu }\partial X^{\nu }\partial
X^{\lambda }e^{ik_{2}\cdot X}e^{ik_{3}\cdot X}e^{ik_{4}\cdot X}>,
\label{22a}
\end{equation}%
\begin{equation}
T_{2}^{(\mu \nu )}=\int \overset{4}{\underset{i=1}{\prod }}%
dx_{i}<e^{ik_{1}\cdot X}\partial ^{2}X^{(\mu }\partial X^{\nu
)}e^{ik_{2}\cdot X}e^{ik_{3}\cdot X}e^{ik_{4}\cdot X}>,  \label{23a}
\end{equation}%
\begin{equation}
T_{2}^{\mu }=1/2\int \overset{4}{\underset{i=1}{\prod }}dx_{i}<e^{ik_{1}%
\cdot X}\partial ^{3}X^{\mu }e^{ik_{2}\cdot X}e^{ik_{3}\cdot
X}e^{ik_{4}\cdot X}>.  \label{24a}
\end{equation}%
The polarization triplet $\{\varepsilon _{(\mu \nu \lambda )},\varepsilon
_{(\lambda \mu )},\varepsilon _{\mu }\}$ and the amplitude triplet $\{T^{\mu
\nu \lambda },T^{(\mu \nu )},T^{\mu }\}$ as given above describe the $\Psi
_{1}$ state. The second state $\Psi _{2}$ is given by%
\begin{equation}
\varepsilon _{\lbrack \mu \nu ]}\alpha _{-1}^{[\mu }\alpha _{-2}^{\nu ]}|0,k>
\label{25a}
\end{equation}%
with gauge conditions%
\begin{equation}
k^{\mu }\varepsilon _{\lbrack \mu \nu ]}=0,\text{ }-k^{2}=4.  \label{26a}
\end{equation}%
The amplitude is given by 
\begin{equation}
T_{2}^{(\mu \nu )}=\int \overset{4}{\underset{i=1}{\prod }}%
dx_{i}<e^{ik_{1}\cdot X}\partial ^{2}X^{[\mu }\partial X^{\nu
]}e^{ik_{2}\cdot X}e^{ik_{3}\cdot X}e^{ik_{4}\cdot X}>.  \label{27a}
\end{equation}%
Equations(22) -(\ref{24a}) and (\ref{27a}) can be calculated by using the
same technique presented in Eq.(\ref{10a}) to be\ 
\begin{eqnarray}
T_{2}^{\mu \nu \lambda } &=&\frac{\Gamma (-\frac{s}{2}-1)\Gamma (-\frac{t}{2}%
-1)}{\Gamma (\frac{u}{2}+2)}  \notag \\
&&\times \{-t/2(t^{2}/4-1)k_{1}^{\mu }k_{1}^{\nu }k_{1}^{\lambda
}+3(s/2+1)t/2(t/2+1)k_{1}^{(\mu }k_{1}^{\nu }k_{3}^{\lambda )}  \notag \\
&&-3s/2(s/2+1)(t/2+1)k_{1}^{(\mu }k_{3}^{\nu }k_{3}^{\lambda
)}+s/2(s^{2}/4-1)k_{3}^{\mu }k_{3}^{\nu }k_{3}^{\lambda }\},  \label{28a}
\end{eqnarray}%
\begin{eqnarray}
T_{2}^{(\mu \nu )}=\frac{\Gamma (-\frac{s}{2}-1)\Gamma (-\frac{t}{2}-1)}{%
\Gamma (\frac{u}{2}+2)} &&\times \{t/2(t^{2}/4-1)k_{1}^{\mu }k_{1}^{\nu
}-(s/2+1)t/2(t/2+1)k_{1}^{(\mu }k_{3}^{\nu )}  \notag \\
&&+s/2(s/2+1)(t/2+1)k_{3}^{(\mu }k_{1}^{\nu )}-s/2(s^{2}/4-1)k_{3}^{\mu
}k_{3}^{\nu }\},  \label{29a}
\end{eqnarray}%
\begin{equation}
T_{2}^{\mu }=\frac{\Gamma (-\frac{s}{2}-1)\Gamma (-\frac{t}{2}-1)}{\Gamma (%
\frac{u}{2}+2)}\times \{s/2(s^{2}/4-1)k_{3}^{\mu }-t/2(t^{2}/4-1)k_{1}^{\mu
}\},  \label{30a}
\end{equation}%
\begin{eqnarray}
T_{2}^{[\mu \nu ]}=\frac{\Gamma (-\frac{s}{2}-1)\Gamma (-\frac{t}{2}-1)}{%
\Gamma (\frac{u}{2}+2)} &&\times \{s/2(s/2+1)(t/2+1)  \notag \\
&&+(s/2+1)t/2(t/2+1)k_{3}^{[\mu }k_{1}^{\nu ]}\}.  \label{31a}
\end{eqnarray}%
The zero-norm states of this mass level can be calculated to be

$C,$ \ \ \ 
\begin{equation*}
\lbrack k_{\lambda }\theta _{\mu \nu }^{\prime }\alpha _{-1}^{\lambda
}\alpha _{-1}^{\mu }\alpha _{-1}^{\nu }+2\theta _{(\mu \nu )}^{\prime
}\alpha _{-1}^{\mu }\alpha _{-2}^{\nu }]|0,k>,
\end{equation*}%
\begin{equation}
\theta _{\mu \nu }^{\prime }=\theta _{\nu \mu }^{\prime },\text{ \ }k^{\mu
}\theta _{\mu \nu }^{\prime }=\eta ^{\mu \nu }\theta _{\mu \nu }^{\prime }=0;
\label{32a}
\end{equation}

$D_{1},$%
\begin{equation*}
\{(5/2k_{\mu }k_{\nu }\theta _{\lambda }^{\prime }+\eta _{\mu \nu }\theta
_{\lambda }^{\prime })\alpha _{-1}^{\nu }\alpha _{-1}^{\mu }\alpha
_{-1}^{\lambda }+9k_{(\mu }\theta _{\nu )}^{\prime }\alpha _{-2}^{\mu
}\alpha _{-1}^{\nu }+6\theta _{\mu }^{\prime }\alpha _{-3}^{\mu }\}|0,k>,
\end{equation*}%
\begin{equation}
k\cdot \theta ^{\prime }=0;  \label{33a}
\end{equation}

$D_{2},$%
\begin{equation*}
\{(1/2k_{\mu }k_{\nu }\theta _{\lambda }+2\eta _{\mu \nu }\theta _{\lambda
})\alpha _{-1}^{\nu }\alpha _{-1}^{\mu }\alpha _{-1}^{\lambda }+9k_{[\mu
}\theta _{\nu ]}\alpha _{-2}^{[\mu }\alpha _{-1}^{\nu ]}-6\theta _{\mu
}\alpha _{-3}^{\mu }\}|0,k>,
\end{equation*}%
\begin{equation}
k\cdot \theta =0;  \label{34a}
\end{equation}

$E,$%
\begin{equation}
\{(17/4k_{\mu }k_{\nu }k_{\lambda }+9/2\eta _{\mu \nu }k_{\lambda })\alpha
_{-1}^{\nu }\alpha _{-1}^{\mu }\alpha _{-1}^{\lambda }+(9\eta _{\mu \nu
}+21k_{\mu }k_{\nu })\alpha _{-1}^{\mu }\alpha _{-2}^{\nu }+25k_{\mu }\alpha
_{-3}^{\mu }\}|0,k>,  \label{35a}
\end{equation}%
where $D_{1}$ and $D_{2}$ states are obtained by symmetrizing and
antisymmetrizing those terms which involve $\alpha _{-1}^{\mu }\alpha
_{-2}^{\nu }$ in the original type I and type II vector zero-norm states.
Ward identities can now be easily written down%
\begin{equation}
k_{\lambda }\theta _{\mu \nu }^{\prime }T_{2}^{(\mu \nu \lambda )}+2\theta
_{\mu \nu }^{\prime }T_{2}^{(\mu \nu )}=0,  \label{36aa}
\end{equation}

\bigskip 
\begin{equation}
(5/2k_{\mu }k_{\nu }\theta _{\lambda }^{\prime }+\eta _{\mu \nu }\theta
_{\lambda }^{\prime })T_{2}^{(\mu \nu \lambda )}+9k_{\mu }\theta _{\nu
}^{\prime }T_{2}^{(\mu \nu )}+6\theta _{\mu }^{\prime }T_{2}^{\mu }=0,
\label{37a}
\end{equation}%
\begin{equation}
(1/2k_{\mu }k_{\nu }\theta _{\lambda }+2\eta _{\mu \nu }\theta _{\lambda
})T_{2}^{(\mu \nu \lambda )}+9k_{\mu }\theta _{\nu }T_{2}^{[\mu \nu
]}-6\theta _{\mu }T_{2}^{\mu }=0,  \label{38a}
\end{equation}%
\begin{equation}
(4/17k_{\mu }k_{\nu }k_{\lambda }+9/2\eta _{\mu \nu }k_{\lambda
})T_{2}^{(\mu \nu \lambda )}+(9\eta _{\mu \nu }+21k_{\mu }k_{\nu
})T_{2}^{(\mu \nu )}+25k_{\mu }T_{2}^{\mu }=0.  \label{39a}
\end{equation}%
Equations (\ref{36aa}) -(\ref{39a}) can be explicitly verified after some
lengthy algebra. One has to check the vanishing of a polynomial of sixth
degree in $s,t$ and $\theta $ in each case. It is now easy to see the
identities (\ref{36aa}), (\ref{37a}) and(\ref{39a}) correspond to $\Psi _{1}$
state, whereas identity (\ref{38a}) generated by the $D_{2}$ zero-norm state
relates amplitudes for $\Psi _{1}$and $\Psi _{2}$ states. Equation (\ref{38a}%
) implies that $\Psi _{1}$and $\Psi _{2}$ indeed form a gauge multiplet.
This is consistent with the result obtained in Ref. \cite{11} from the
worldsheet $\sigma $-model point of view. Note that one has to consider the $%
amplitude$ $triplet$ of $\Psi _{1},\{T_{2}^{(\mu \nu \lambda )},T_{2}^{(\mu
\nu )},T_{2}^{\mu }\}$, in order to see apparently the inter-particle
symmetry. The usual\textquotedblleft gauge choice\textquotedblright for $%
\Psi _{1},\{\varepsilon _{(\mu \nu \lambda )},T_{2}^{(\mu \nu \lambda )}\}$
with traceless and transverse gauge conditions, will $hide$ this interesting
\textquotedblleft hidden stringy symmetry \textquotedblright .

We can now derive the closed string Ward identities by using the
relationship between closed and open string amplitudes \cite{14} 
\begin{equation}
A_{closed}=-\pi \kappa ^{2}\sin (\pi k_{2}\cdot k_{3})A_{open}(s,t)\overline{%
A}_{open}(t,u).  \label{40a}
\end{equation}%
For example, the Ward identity corresponding to $D_{2}\otimes D_{2}^{\prime }
$ zero-norm state is%
\begin{eqnarray*}
&&(1/4k_{\mu }k_{\nu }k_{\alpha }k_{\beta }\theta _{\lambda \gamma }+\eta
_{\mu \nu }k_{\alpha }k_{\beta }\theta _{\lambda \gamma }+\eta _{\alpha
\beta }k_{\mu }k_{\nu }\theta _{\lambda \gamma }+4\eta _{\mu \nu }\eta
_{\alpha \beta }\theta _{\lambda \gamma })T_{2}^{(\mu \nu \lambda )(\alpha
\beta \gamma )} \\
&&+(9/2k_{\mu }k_{\nu }k_{\alpha }\theta _{\lambda \beta }+18\eta _{\mu \nu
}k_{\alpha }\theta _{\lambda \beta })T_{2}^{(\mu \nu \lambda )[\alpha \beta
]} \\
&&+(9/2k_{\alpha }k_{\beta }k_{\mu }\theta _{\gamma \nu }+18\eta _{\alpha
\beta }k_{\mu }\theta _{\gamma \nu })T_{2}^{[\mu \nu ](\alpha \beta \gamma )}
\\
&&+81k_{\mu }k_{\alpha }\theta _{\nu \beta }T_{2}^{[\mu \nu ][\alpha \beta
]}-(3k_{\mu }k_{\nu }\theta _{\lambda \alpha }+12\eta _{\mu \nu }\theta
_{\lambda \alpha })T_{2}^{(\mu \nu \lambda )\alpha } \\
&&-(3k_{\alpha }k_{\beta }\theta _{\gamma \lambda }+12\eta _{\alpha \beta
}\theta _{\gamma \nu })T_{2}^{\lambda (\alpha \beta \gamma )}-54k_{\mu
}\theta _{\nu \alpha }T_{2}^{[\mu \nu ]\alpha }
\end{eqnarray*}%
\begin{equation}
-54k_{\alpha }\theta _{\beta \mu }T_{2}^{\mu \lbrack \alpha \beta
]}+36\theta _{\mu \lambda }T_{2}^{\mu \lambda }=0,  \label{41a}
\end{equation}%
where $\theta _{\mu \nu }$ is a constant tensor with $k^{\mu }\theta _{\mu
\nu }=k^{\nu }\theta _{\mu \nu }=0.$ In Eq. (\ref{41a}) the definition of,
e.g., $T_{2}^{(\mu \nu \lambda )[\alpha \beta ]}$ is%
\begin{equation}
T_{2}^{(\mu \nu \lambda )[\alpha \beta ]}=\int \overset{4}{\underset{i=1}{%
\prod }}d^{2}z_{i}<e^{ik_{1}\cdot X}\partial X^{(\mu }\partial X^{\nu
}\partial X^{\lambda )}\overline{\partial ^{2}}X^{[\alpha }\overline{%
\partial }X^{\beta ]}e^{ik_{2}\cdot X}e^{ik_{3}\cdot X}e^{ik_{4}\cdot X}>,
\label{42a}
\end{equation}%
others can be similarly written down. One can also write down the symmetry
transformation in the first order weak field approximation for Eq.(\ref{41a}%
). Let $M_{\mu \nu \lambda ,\alpha \beta \gamma }(X),E_{\mu \nu ,\alpha
\beta \gamma }(X),G_{\mu \nu \lambda ,\alpha \beta }$ and $A_{\mu \nu
,\alpha \beta }(X)$ be the four physical propagating background fields of
the second massive level $(\mu \nu \lambda \equiv (\mu \nu \lambda )$ and $%
\mu \nu \equiv \lbrack \mu \nu ])$, the inter-particle gauge transformation
is ( replace $k_{\mu }$ by $\partial _{\mu }$ and $\eta _{\mu \nu }$ by $%
-\eta _{\mu \nu }$ in Eq.(\ref{41a}) according to our convention in Ref. 
\cite{7})%
\begin{equation*}
\delta M_{\mu \nu \lambda ,\alpha \beta \gamma }=1/4\partial _{(\mu
}\partial _{\nu }\partial _{(\alpha }\partial _{\beta }\theta _{\lambda
)\gamma )}-\eta _{(\mu \nu }\partial _{(\alpha }\partial _{\beta }\theta
_{\lambda )\gamma )}-\eta _{(\alpha \beta }\partial _{(\mu }\partial _{\nu
}\theta _{\lambda )\gamma }+4\eta _{(\mu \nu }\eta _{(\alpha \beta }\theta
_{\lambda )\gamma )},
\end{equation*}%
\begin{equation*}
\delta E_{\mu \nu ,\alpha \beta \gamma }=9/2\partial _{(\alpha }\partial
_{\beta }\partial _{\lbrack \mu }\theta _{\gamma )\nu ]}-18\eta _{(\alpha
\beta }\partial _{\lbrack \mu }\theta _{\gamma )\nu ]},
\end{equation*}%
\begin{equation*}
\delta G_{\mu \nu \lambda ,\alpha \beta }=9/2\partial _{(\mu }\partial _{\nu
}\partial _{\lbrack \alpha }\theta _{\lambda )\beta ]}-18\eta _{(\mu \nu
}\partial _{\lbrack \alpha }\theta _{\lambda )\beta ]},
\end{equation*}%
\begin{equation}
\delta A_{\mu \nu ,\alpha \beta }=81\partial _{\lbrack \mu }\partial
_{\lbrack \alpha }\theta _{\nu ]\beta },  \label{43a}
\end{equation}%
where $\partial ^{\mu }\theta _{\mu \nu }(X)=$ $\partial ^{\nu }\theta _{\mu
\nu }(X)=0$ and $(\Box -4)\theta _{\mu \nu }(X)=0$, the other five
background fields are just gauge artifacts according to Eq.(\ref{21aa}).
Equation (\ref{43a}) is consistent with the result we obtained from
generalized $\sigma $-model point of view in Ref \cite{11}. However, the
method used in this paper can be easily generalized to superstring massive
cases, which turn out to be too messy to obtain from $\sigma $-model
approach.\cite{13}

We have considered only four-point amplitudes which contain only one
non-tachyon particle. In general, we can consider $n$-point amplitudes
consisting of any kind of massive states. This will give us all kinds of
Ward identities including those which are associated with inter-mass level
symmetries. \cite{16} Moreover, our construction has interesting implication
on low dimensional string theory. Work in this direction is currently in
progressed.

I would like to thank Professor H. Kawai for useful suggestions. I thank
Chuan-Tsung Chan for many corrections of coefficients in the paper for me.
This work was supported in part by the National Science Council of R.O.C
under contract \#NSC-81-0208-M-009-505.

\bigskip

%-----------------------------------------------------------------------------
%\bibliographystyle{amsplain}

%-----------------------------------------------------------------------------

\end{document}